\documentclass[fleqn,10pt]{wlscirep}
\usepackage[utf8]{inputenc}
\usepackage[T1]{fontenc}
\usepackage{lineno}
\usepackage{caption}

\title{Age structure alters optimal hospital allocation for reducing tuberculosis fatalities in South Korea}

\author[1]{Yongsung Kwon}
\author[2*]{Deok-Sun Lee}
\author[3*]{Mi Jin Lee}
\author[1,4,*]{Seung-Woo Son}
\affil[1]{Department of Applied Artificial Intelligence, Hanyang University, Ansan 15588, Korea}
\affil[2]{School of Computational Sciences, Korea Institute for Advanced Study, Seoul 02455, Korea}
\affil[3]{Department of Physics, Pusan National University, Busan 46241, Korea}
\affil[4]{Department of Applied Physics, Hanyang University, Ansan 15588, Korea}

\affil[*]{Corresponding Author(s): Deok-Sun Lee (deoksunlee@kias.re.kr), Mi Jin Lee (mijinlee@pusan.ac.kr), Seung-Woo Son (sonswoo@hanyang.ac.kr)}

\begin{abstract}
Optimal healthcare allocation requires accounting for spatial heterogeneity in disease burden, while demographic heterogeneity may further alter local vulnerability when outcomes vary strongly with age. Tuberculosis (TB) provides a useful case because treatment requires sustained access to care and fatality risk rises markedly with age. In South Korea, TB incidence remains relatively high, while regional incidence, mortality, and hospital data are systematically recorded, allowing these effects to be examined empirically. Finer spatial resolution provides more local observations, helping reveal regional heterogeneity and spatial patterns. However, due to privacy concerns, official district-level TB statistics are provided only in age-aggregated form, whereas age-specific statistics are available at province level. We therefore propose a method to reconstruct age-resolved TB cases and fatalities across 228 districts for 2014--2022 by combining province-level age distributions with district-level totals, using the finest available age and spatial information. Building on an existing hospital-allocation framework, we incorporate age-dependent vulnerability into the fatality-minimization objective. We find that age-aware and age-agnostic optimizations yield similar total minimized fatalities but distinct district-level allocations. The oldest age group's rescaled patient density is strongly associated with this difference and the direction of hospital redistribution. Age-weighting schemes clarify how countervailing contributions across age groups affect optimization.
\end{abstract}

\begin{document}

\flushbottom
\maketitle

\thispagestyle{empty}

\section*{Introduction}

Tuberculosis (TB) remains one of the leading causes of infectious disease mortality worldwide, with an estimated 1.3 million deaths annually~\cite{WHO2023, glaziou2015global}. Despite long-term global efforts to control the disease, substantial regional disparities in TB incidence and outcomes persist~\cite{renner2024hospitals, chung2024access}. Within the Organization for Economic Co-operation and Development (OECD), South Korea presents a notable case: it consistently reports over 10\,000 new TB cases annually and ranks second in TB incidence and fifth in TB-related mortality among OECD countries~\cite{lee2025global}. Because effective TB treatment requires early diagnosis and repeated visits to secondary care facilities over extended periods~\cite{cdc2013tbcurriculum, hopewell2006international}, the spatial distribution of hospitals directly shapes disease outcomes. Combined with a fully digitized surveillance system that reports TB patients, fatalities, and healthcare infrastructure at the administrative district level~\cite{KDCA, KOSIS}, this makes South Korea a natural setting in which to ask how a fixed stock of healthcare infrastructure should be distributed in order to minimize fatalities.

The spatial distribution of public infrastructure has been studied as a generic feature of urban systems, where facility density scales with population density~\cite{um2009scaling, stephan1977territorial, gusein1982bunge, gastner2006optimal, kim2012internet}. In the context of healthcare, such spatial organization influences not only average accessibility but also the robustness of service provision under uneven demand or localized constraints, as observed in other infrastructure systems~\cite{lee2017spatial, wuellner2010resilience, lee2022spatial, kwon2025quantifying}. Within this line of work, a previous study formulated TB fatalities as an explicit function of areal hospital density using a random-walk model with traps~\cite{lee2020optimizing, hughes1996random}, and showed that reallocating a fixed number of hospitals across districts can substantially reduce nationwide fatalities. The appeal of this formulation is that it converts a policy question---where hospitals should be located---into a well-posed optimization problem whose parameters are fixed entirely by empirical data.

Such frameworks, however, treat all patients as equally vulnerable. TB severity depends strongly on age~\cite{lee2025global, Rajagopalan2001}: older individuals are more likely to experience fatal outcomes and may face structural barriers to early diagnosis or continuous treatment~\cite{hopewell2006international, cdc2013tbcurriculum}. In South Korea, TB fatalities are heavily concentrated among individuals aged 70 years and older [Fig.~\ref{fig:phi_age_map}(a)], whereas the general population is predominantly concentrated in middle-aged groups, indicating a pronounced mismatch between demographic structure and TB mortality risk. Age composition is itself spatially heterogeneous: metropolitan districts have markedly younger populations than non-metropolitan ones [Figs.~\ref{fig:phi_age_map}(b) and~\ref{fig:phi_age_map}(c)]. An age-agnostic optimization therefore suppresses two sources of variation that need not act in the same direction, and it is not obvious a priori whether accounting for them changes the resulting spatial allocation, changes only the achievable reduction in fatalities, or changes both. This is the question we address here.

Answering this question requires TB data that are jointly disaggregated by age and by administrative district. Such data are not publicly released in South Korea: age-specific counts are available only at the province level (17 units), while district-level counts (228 units) are age-aggregated, a restriction imposed by privacy regulations~\cite{KDCA}. We therefore reconstruct age-resolved district-level counts of TB cases and fatalities for the period 2014--2022 by combining province-level age distributions with district-level totals, an upscaling procedure that introduces age resolution while preserving spatial fidelity and exact consistency with the published marginals. The reconstruction uses only publicly available aggregated statistics and does not involve individual-level records. The procedure assumes the age composition of TB cases and fatalities to be homogeneous within each province; we return to the implications of this assumption in the Discussion.

Using the reconstructed data, we extend the fatality-minimization framework in two ways. First, the characteristic hospital density is made age-dependent, so that each age group within a district experiences its own effective accessibility. Second, a linear age-weighting parameter is introduced to vary the relative contributions of different age groups to the objective function under a fixed total number of hospitals. Because a closed-form solution is no longer available in this age-resolved formulation, optimal configurations are obtained numerically by zero-temperature Monte Carlo relocation. We show that the rescaled patient density of the oldest age group is associated with both whether a district gains or loses hospitals after optimization and the difference between the age-aware and age-agnostic optima. We further show that under different age-weighting schemes, district-level changes in hospital allocation do not necessarily correspond directly to changes in the weighted fatality objective: the net outcome reflects countervailing contributions from different age groups. These results show that age-resolved contributions can reveal district-level differences that are not captured by aggregate optimization outcomes alone.

\section*{Background and Motivation}
\label{sec:background}

To optimize hospital density for minimizing TB fatalities, a previous study~\cite{lee2020optimizing} formulated the TB fatalities $E$ as a function of hospital density while accounting for human mobility, based on a random-walk model with traps~\cite{hughes1996random}, as follows:
\begin{equation}
    E({\vec{\eta}}) = \sum_s N_s \phi_s(\eta_s) = \sum_s N_s \exp(-\eta_s/\tilde{\eta}_s),
    \label{eq:energy_previous}
\end{equation}
where the hospital density $\eta_s$ and the fatality rate $\phi_s$ for district $s$ are defined as $\eta_s \equiv H_s/A_s$ and $\phi_s \equiv D_s/N_s$, respectively. Following the previous study~\cite{lee2020optimizing}, the annual ratio $D_s/N_s$ is used here as an empirical approximation of the TB fatality rate. Here, $H_s$ is the number of hospitals, $A_s$ is the district area, and $D_s$ and $N_s$ are the numbers of TB-related deaths and newly reported TB cases, respectively, in a given year. The hospital configuration across all districts is denoted by $\vec{\eta}=(\eta_1, \eta_2, \cdots, \eta_s, \cdots)$. The characteristic hospital density $\tilde\eta_s$, which reflects the medical and infrastructural environment, can be empirically determined by equating the observed fatality rate ($\phi_s\equiv D_s/N_s$) to the modeled form $\phi_s=\exp(-\eta_s/\tilde\eta_s)$, yielding $\tilde\eta_s=\eta_s/\log (N_s/D_s)$, as reported in the previous study~\cite{lee2020optimizing}. Preserving the total number of hospitals, the optimal hospital density of district $s$ is theoretically obtained as $\eta_s^{\rm(opt)}=\tilde{\eta}_s \log \left( {\rho_s} / {(z\tilde{\eta}_s)}\right)$, where $\rho_s= {N_s}/{A_s}$ denotes the patient density and $z$ is a Lagrange multiplier. The previous study further demonstrated that the ratio ${\eta^{\rm (opt)}}/{\eta}$ between the optimal and current hospital densities strongly depends on the characteristic patient density ${\rho_s}/{\tilde{\eta}_s}$~\cite{lee2020optimizing}.

\begin{figure}[ht!]
\centering
\includegraphics[width=1\columnwidth]{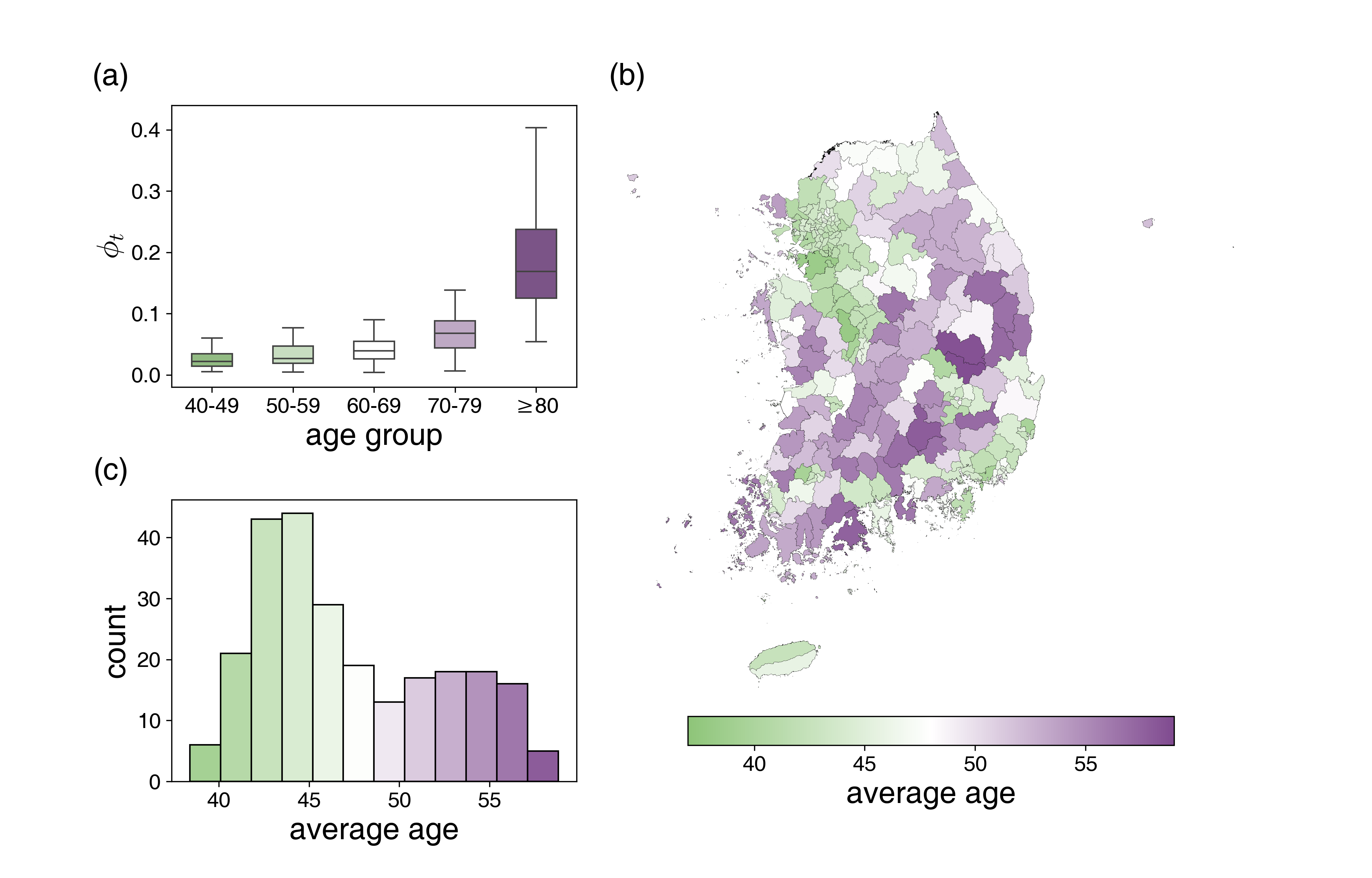}
\caption{(a) Quartile plots of the fatality rate $\phi_t$ by age group $t$ across 16 provinces in 2022 (excluding Sejong-si due to the absence of reported fatalities). (b) Average age of the general population across the districts. Metropolitan districts tend to have younger populations, while non-metropolitan districts are characterized by older populations. (c) The distribution of average age displays a heterogeneous pattern in South Korea, indicating an overconcentration of younger individuals in metropolitan areas.}
\label{fig:phi_age_map}
\end{figure}

This modeling framework assumes that the fatality rate $\phi$ depends solely on the characteristics of each district and therefore does not capture the well-known dependence of TB fatality on patient age~\cite{WHO2023,Rajagopalan2001}. This simplification is primarily due to data limitations: as of 2022, South Korea comprises 228 ``si-gun-gu'' districts (municipal government level, higher-resolution administrative units), which belong to 17 ``si-do'' provinces (regional local government level, lower-resolution units). Each province includes as few as 2 districts (Jeju Special Self-Governing Province) and as many as 31 districts (Gyeonggi-do, part of the Seoul metropolitan area). TB data at the district level are more suitable for resolving spatial heterogeneity and systematic regional patterns, but age-disaggregated TB data are only available at the province level due to privacy concerns. Figure~\ref{fig:phi_age_map}(a) illustrates the age-dependent TB fatality rates for the year 2022. Data provided by the Korea Disease Control and Prevention Agency (KDCA)~\cite{KDCA} and Statistics Korea (KOSIS)~\cite{KOSIS} include the number $N_t$ of newly reported TB cases and the number $D_t$ of TB-related deaths for each age group $t$, aggregated at the province level. As shown in Fig.~\ref{fig:phi_age_map}(a), the fatality rate $\phi_t = D_t/N_t$ exhibits substantial variation across age groups, with median values ranging from 0.0219 to 0.1690. Age groups under 40 years are not shown, as they exhibit negligible fatality rates.

In addition to age-related variability, the wide range of fatality rates observed within each age group suggests the presence of substantial spatial heterogeneity. While age-disaggregated TB data are not available at the district level, spatial differences can be indirectly inferred from the age composition of the general population. As shown in Fig.~\ref{fig:phi_age_map}(b), metropolitan districts, including the capital city Seoul, tend to have younger average populations than non-metropolitan districts, resulting in a skewed age distribution across regions. This demographic contrast implies that the age composition of TB patients is also likely to vary spatially at the district level. These considerations motivate the need for TB data that jointly capture age structure and spatial resolution, which we address by constructing a higher-resolution, age-inclusive dataset and extending the fatality rate formulation accordingly.

\section*{Methods}
\begin{figure}
    \centering
    \includegraphics[width=1\linewidth]{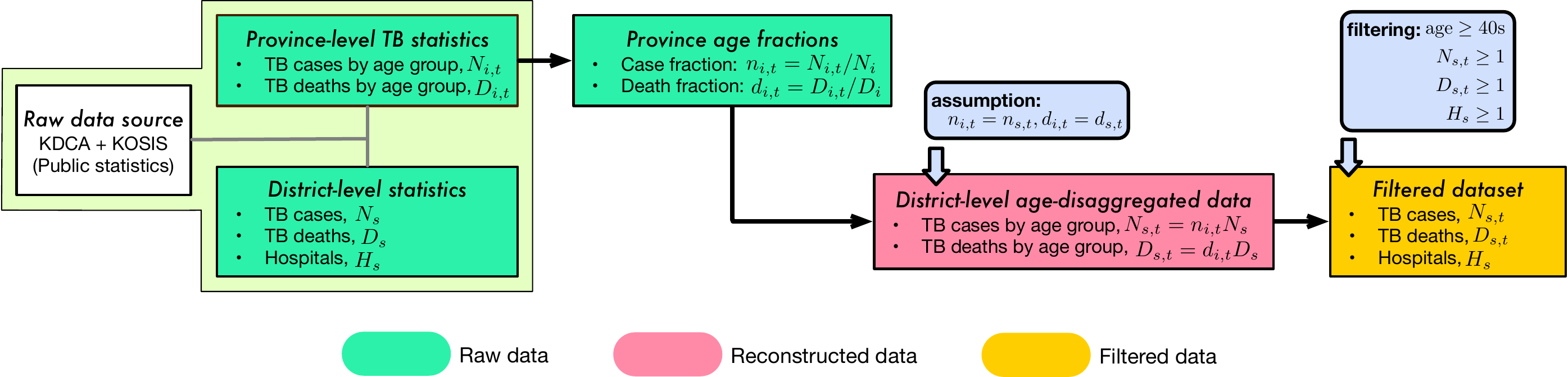}
\caption{Flowchart of the data construction process. Starting from the raw source data (green), we reconstruct the district-level age-disaggregated quantities (pink). By construction, the reconstructed and raw quantities satisfy the relations $N_{s}=\sum_t N_{s,t}$ and $D_{s}=\sum_t D_{s,t}$. The raw source data shown in green correspond to the records summarized in Table~\ref{tab:data_info_sido}. For subsequent analyses, the reconstructed dataset is further filtered using empirical criteria. The final dataset (yellow) contains districts $s$ that satisfy $H_s \geq 1$ and $N_{s,t}, D_{s,t} \geq 1$ for all age groups $t$ of 40 years and older, corresponding to Table~\ref{tab:data_info}.}
    \label{fig:flowchart}
\end{figure}

We reconstruct age-resolved district-level TB counts by combining province-level age distributions with district-level totals and filter the reconstructed data for the optimization analysis. The overall procedure is summarized in Fig.~\ref{fig:flowchart}.

\subsection*{Raw data sources and spatial resolution} \label{sec:data}

\begin{figure}[ht]
\centering
\includegraphics[width=1\columnwidth]{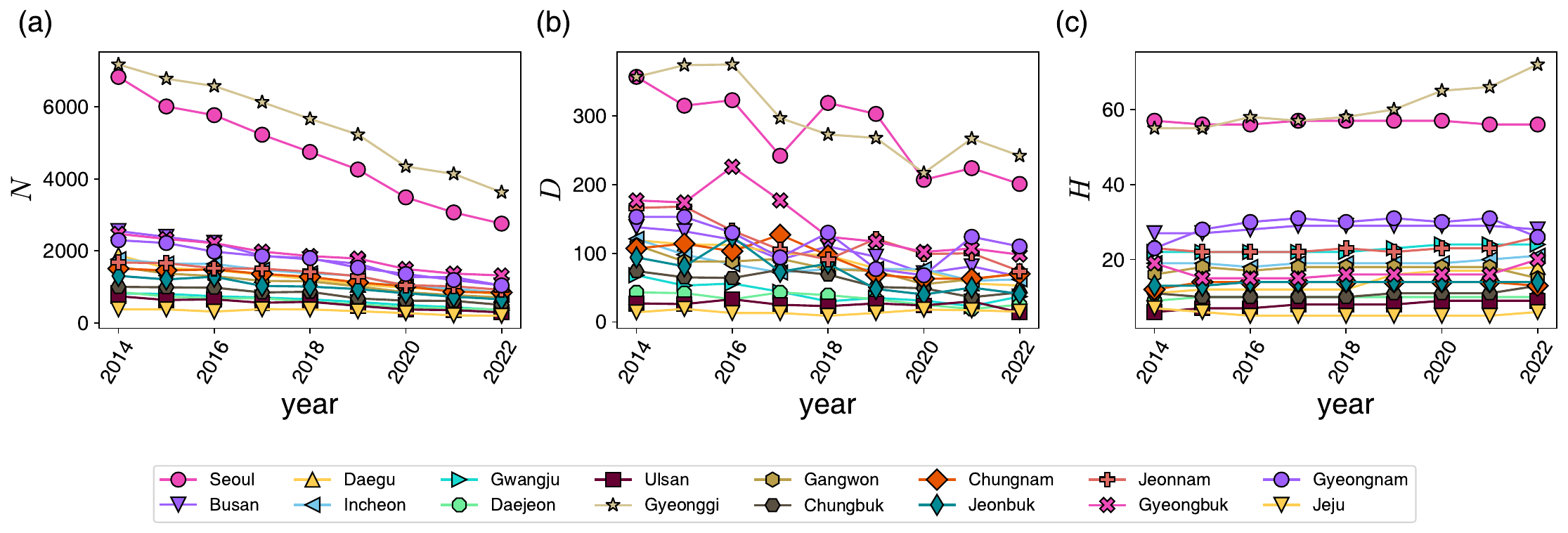}
\caption{Temporal patterns of the raw data for (a) the number $N$ of newly reported TB cases, (b) the number  $D$ of TB-related deaths, and (c) the number $H$ of hospitals. The number of new TB cases and deaths is showing a declining trend. The number of secondary care hospitals is slightly increasing in the Gyeonggi-do region. The 16 provinces are represented by distinct symbols. 
}
\label{fig:hND}
\end{figure}

We utilize publicly available datasets reported annually from 2014 to 2022, provided by KDCA~\cite{KDCA} and KOSIS~\cite{KOSIS}. This study requires age-disaggregated demographic and TB-related information at the district level, the higher-resolution administrative unit. Over this nine-year period, data have been collected for 17 provinces and 228 districts. In contrast to demographic information, such as the number $N$ of newly reported TB cases and the number $D$ of TB-related deaths by age group, non-demographic information, including the number $H$ of hospitals, is available at the district scale. Here, ``hospital'' refers to a secondary care hospital equipped to provide appropriate treatment for TB. The annual totals of each dataset are summarized in Table~\ref{tab:data_info_sido}, and a subset of the data is visualized regionally in Fig.~\ref{fig:hND}.

\begin{table}[h!]
\centering
\setlength{\tabcolsep}{10pt}
\begin{tabular}{|c|c|c||c|}
\hline
Year  & $N$ & $D$ & $H$ \\
\hline
2014 & 34 869 & 2 136 & 330 \\
\hline
2015 & 32 182 & 2 018 & 335 \\
\hline
2016 & 30 892 & 2 020 & 339 \\
\hline
2017 & 28 161 & 1 678 & 344 \\
\hline
2018 & 26 433 & 1 657 & 346 \\
\hline
2019 & 23 821 & 1 492 & 354 \\
\hline
2020 & 19 933 & 1 222 & 360 \\
\hline
2021 & 18 335 & 1 324 & 364 \\
\hline
2022 & 16 264 & 1 223 & 374 \\
\hline
\end{tabular}
\caption{\label{tab:data_info_sido}
Total annual records of the TB source data. Age-disaggregated counts of newly reported TB cases ($N$) and TB-related deaths ($D$) are available only at the province level due to privacy concerns, whereas age-aggregated district-level counts are also available for spatial reconstruction. The number $H$ of hospitals is provided at the district level. Here, ``hospital'' refers to a secondary care hospital equipped to provide appropriate treatment for TB.
}
\end{table}

Figure~\ref{fig:hND} shows the temporal patterns of these data for each province separately. Seoul-si and Gyeonggi-do exhibit markedly higher values across all measured quantities than other provinces, reflecting their combined population of nearly half of South Korea’s total population (approximately 22 million out of 50 million). The number $N$ of newly reported TB cases shows a consistent downward trend across all provinces [Fig.~\ref{fig:hND}(a)], and the number $D$ of TB-related deaths also decreases with small fluctuations [Fig.~\ref{fig:hND}(b)], indicating steady nationwide progress toward TB control. The number $H$ of hospitals remains nearly unchanged in most provinces, except for Gyeonggi-do. As a major metropolitan region surrounding the capital city Seoul, Gyeonggi-do has continued to experience population growth driven by suburbanization. This population increase has been accompanied by a growing demand for infrastructure, including hospitals.

\begin{figure}[t]
\centering
\includegraphics[width=1\columnwidth]{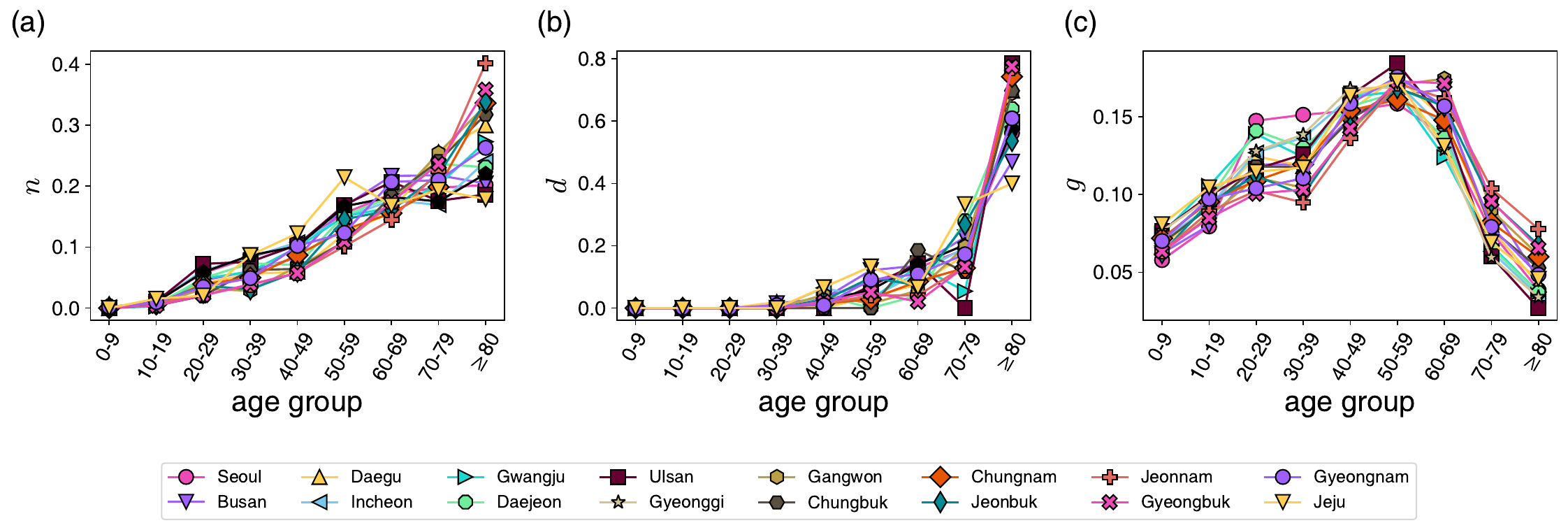}
\caption{Regional and age-grouped tendency of (a) TB patients fraction $n$, (b)  TB deaths fraction $d$, and, as a comparison, (c) the general population demographic fraction $g$ for the respective province in 2022. Both the proportion of new TB cases and TB-related deaths tend to increase with age. The 16 provinces are represented by distinct symbols.}
\label{fig:age_distribution}
\end{figure}

To examine the age-group composition of TB cases and fatalities, we measure the fractions $n_{i,t}$ and $d_{i,t}$ of individuals in age group $t$ among newly reported TB cases and TB-related deaths in province $i$, respectively, as
\begin{align}
    n_{i,t} = {N_{i,t}}/{N_i}, ~~~~~~~
    d_{i,t} = {D_{i,t}}/{D_i}, \label{eq:ndit}
\end{align}
where $N_i=\sum_{t} N_{i,t}$ and $D_i=\sum_{t} D_{i,t}$. Ages are categorized into ten-year groupings. Similarly, the quantities $N_t$ and $D_t$, used to compute the age-dependent fatality rate $\phi_t$ shown in Fig.~\ref{fig:phi_age_map}(a), are calculated as $N_t=\sum_i N_{i,t}$ and $D_t=\sum_i D_{i,t}$, respectively. The resulting age-group fractions are plotted in Fig.~\ref{fig:age_distribution}. The TB-related fractions $n$ and $d$, shown in Figs.~\ref{fig:age_distribution}(a) and~\ref{fig:age_distribution}(b), increase monotonically with age. Notably, age dependence is more pronounced for TB-related deaths than for newly reported cases. Moreover, TB deaths in Fig.~\ref{fig:age_distribution}(b) contrast with the age distribution of the general population shown in Fig.~\ref{fig:age_distribution}(c): although the population in South Korea is primarily concentrated in the 40--60 year age range, TB-related deaths are disproportionately concentrated among individuals aged 70 years and older. These observations indicate that the age distribution of TB fatalities deviates systematically from the underlying population structure, emphasizing the need to explicitly account for age-dependent effects in the subsequent analysis.

\subsection*{Upscaling TB demographic data to high spatial resolution}
\label{sec:generation}

The raw demographic data related to TB described in the previous subsection are officially available in an age-disaggregated format only at the provincial level. To enable a more fine-grained spatial analysis, we develop a method to reconstruct TB case and fatality counts at the district level. Specifically, we assume that the age distributions observed at the provincial level [Figs.~\ref{fig:age_distribution}(a) and~\ref{fig:age_distribution}(b)] also apply to the districts within each province. Then, we can estimate the numbers of TB cases and deaths in district $s$, belonging to province $i$, for age group $t$ using Eq.~(\ref{eq:ndit}) as
\begin{align}
    N_{s,t} \equiv n_{i,t} N_{s}, ~~~~~
    D_{s,t} \equiv d_{i,t} D_{s}, \label{eq:ndst}
\end{align}
where $N_{s}$ and $D_{s}$ denote the age-aggregated numbers of newly reported TB cases and TB-related deaths in district $s$, respectively, that are available in KDCA and KOSIS. The reconstructed data satisfy $\sum^{\prime}_{s} N_{s,t} = N_{i,t}$ and $\sum^{\prime}_{s} D_{s,t} = D_{i,t}$, where the primed summation runs over districts $s$ belonging to province $i$. In this reconstruction framework, the age-group fractions of both TB cases and fatalities are assumed to be homogeneous within each province, such that the province-level fractions $n_{i,t}$ and $d_{i,t}$ are applied to all districts $s \in i$.

\begin{table}[ht]
\centering
\setlength{\tabcolsep}{10pt}
\begin{tabular}{|c|c|c|c||c|}
\hline
Year  & $S$ & $N$ & $D$ & $H$ \\
\hline
2014 & 143 & 32 322 & 1 718 & 328 \\
\hline
2015 & 137 & 21 683 & 1 296 & 312 \\
\hline
2016 & 132 & 20 207 & 1 255 & 306 \\
\hline
2017 & 143 & 19 427 & 1 110 & 330 \\
\hline
2018 & 142 & 18 145 & 1 157 & 334 \\
\hline
2019 & 119 & 14 318 & 901 & 285 \\
\hline
2020 & 116 & 12 079 & 736 & 276 \\
\hline
2021 & 113 & 10 900 & 749 & 262 \\
\hline
2022 & 124 & 11 789 & 886 & 314 \\
\hline
\end{tabular}
\caption{Summary of annual TB data at the district level available from 2014 to 2022. For each year, the table lists the number $S$ of districts considered, the number $N$ of newly reported TB cases across those districts, the number $D$ of TB-related deaths, and the number $H$ of hospitals. The age-resolved counts used in the subsequent analysis are reconstructed from province-level age distributions using Eq.~(\ref{eq:ndst}). All included districts satisfy the conditions of having at least one newly reported TB case ($N \geq 1$), at least one TB-related death among individuals aged 40 years and older ($D \geq 1$), and at least one hospital ($H \geq 1$).
}
\label{tab:data_info}
\end{table}

To implement the optimal distribution of hospitals based on the random walk model as in the previous study~\cite{lee2020optimizing}, the numbers of TB patients, TB-related deaths, and hospitals of each district are used. We extend this formulation to incorporate age effects using Eq.~(\ref{eq:ndst}). In addition, as mentioned earlier and shown in Fig.~\ref{fig:age_distribution}(b), the number of TB-related deaths is extremely low (nearly zero) for age groups under 40. Therefore, we exclude age groups under 40, resulting in five age groups being considered: $t\in \{1, 2, 3, 4, 5\}$ (e.g., $t=1$ denotes the age group ``$40$--$49$''). In summary, the subset of data that satisfies these requirements is extracted from the full set of districts and is presented in Table~\ref{tab:data_info}. The number of districts included in the analysis ranges from 113 to 143, out of a total of 228 districts.

\section*{Results}
\subsection*{Modeling the TB fatality rate and objective function}

Using upscaled TB data in Eq.~(\ref{eq:ndst}), we aim to minimize the total number of TB-related deaths by reallocating hospitals across districts while accounting for age-specific effects. To this end, we reformulate the objective function in Eq.~(\ref{eq:energy_previous}) by decomposing total fatalities into age-group components. Specifically, the original fatality rate $\phi_s$ is extended to $\phi_{s,t}\equiv D_{s,t}/N_{s,t}$, representing the fatality rate for age group $t$ in district $s$. As stated in Eq.~(\ref{eq:energy_previous}), the fatality rate was modeled as $\phi_s(\eta_s) = \exp(-\eta_s / \tilde{\eta}_s)$, where $\eta_s$ denotes the areal density of hospitals and $\tilde{\eta}_s$ represents a district-specific characteristic density. In our age-disaggregated formulation, individuals in district $s$ access a shared set of $H_s$ hospitals regardless of age, implying that the hospital density remains $\eta_{s,t} = \eta_s$. However, hospital accessibility may vary across age groups (e.g., due to differences in mobility and activity), resulting in an age-dependent characteristic density $\tilde{\eta}_{s,t}$. Accordingly, the age-specific fatality rate is modeled as
\begin{equation}
\phi_{s,t} = \exp\left(- {\eta_s} / {\tilde{\eta}_{s,t}}\right),
\label{eq:phi_st}
\end{equation}
with the characteristic density empirically estimated as
\begin{equation}
\tilde{\eta}_{s,t} = \frac{\eta_s}{-\log \phi_{s,t}} = \frac{H_s/A_s}{\log (N_{s,t} / D_{s,t})},
\label{eq:tilde_st}
\end{equation}
using the observed values of $N_{s,t}$, $D_{s,t}$, and $H_s$ summarized in Table~\ref{tab:data_info}. Because $\tilde{\eta}_{s,t}$ sets the decay scale in Eq.~(\ref{eq:phi_st}), a larger $\tilde{\eta}_{s,t}$ corresponds to a slower decrease of the modeled fatality rate with increasing hospital density. Thus, achieving the same proportional reduction in $\phi_{s,t}$ requires a larger increase in hospital density when $\tilde{\eta}_{s,t}$ is larger.

Using the model defined in Eq.~(\ref{eq:phi_st}), the total number of fatalities, which serves as the objective function to be minimized, can be written as a function of hospital densities:
\begin{equation}
E_{\rm fatalities}(\vec{\eta}) = \sum_{s,t} D_{s,t} = \sum_{s,t} N_{s,t} \phi_{s,t} = \sum_{s,t} N_{s,t} \exp\left(- {\eta_s} / {\tilde{\eta}_{s,t}}\right).
\label{eq:energy_st}
\end{equation}
We vary $\vec{\eta}$ to minimize $E_{\rm fatalities}(\vec{\eta})$ while the total number $H$ of hospitals is preserved. The optimal hospital density $\eta_s^{\rm (opt)}$ that minimizes $E_{\rm fatalities}(\vec{\eta})$ can, in principle, be obtained analytically using the Lagrange multiplier method. However, when age groups $t$ are taken into account in Eq.~(\ref{eq:energy_st}), a closed-form solution is not available. Therefore, we perform a numerical optimization to determine the optimized hospital configuration vector $\vec{\eta}^{\rm (opt)} = \left(\eta_1^{\rm (opt)}, \eta_2^{\rm (opt)}, \cdots, \eta_s^{\rm (opt)}, \cdots\right)$ using a zero-temperature Monte Carlo approach. A randomly selected pair $(u, v)$ of districts attempts to exchange a small amount of hospital allocation, denoted by $\Delta H$, through the updates $H_u \to H_u - \Delta H$ and $H_v \to H_v + \Delta H$. If this relocation results in a decrease in the total fatalities $E_{\rm fatalities}$, the new configuration is accepted. This process is repeated until $E_{\rm fatalities}$ converges to a minimum value, denoted by $E^{\rm (min)}_{\rm fatalities}$. The optimized density configuration is defined as the one that minimizes the objective function, i.e.,
\begin{equation}
E^{\rm (min)}_{\rm fatalities} \equiv E\left(\vec{\eta}^{\rm (opt)} \right).
\label{eq:energy_min}
\end{equation}
Following Ref.~\cite{lee2020optimizing}, $H_s$ is treated as a continuous variable during optimization; thus, $\Delta H$ represents the step size of the continuous relaxation rather than the literal relocation of a fractional physical hospital. Here we use $\Delta H = 0.01$ as the numerical step size.

\begin{figure}[t!]
\centering
\includegraphics[width=1\columnwidth]{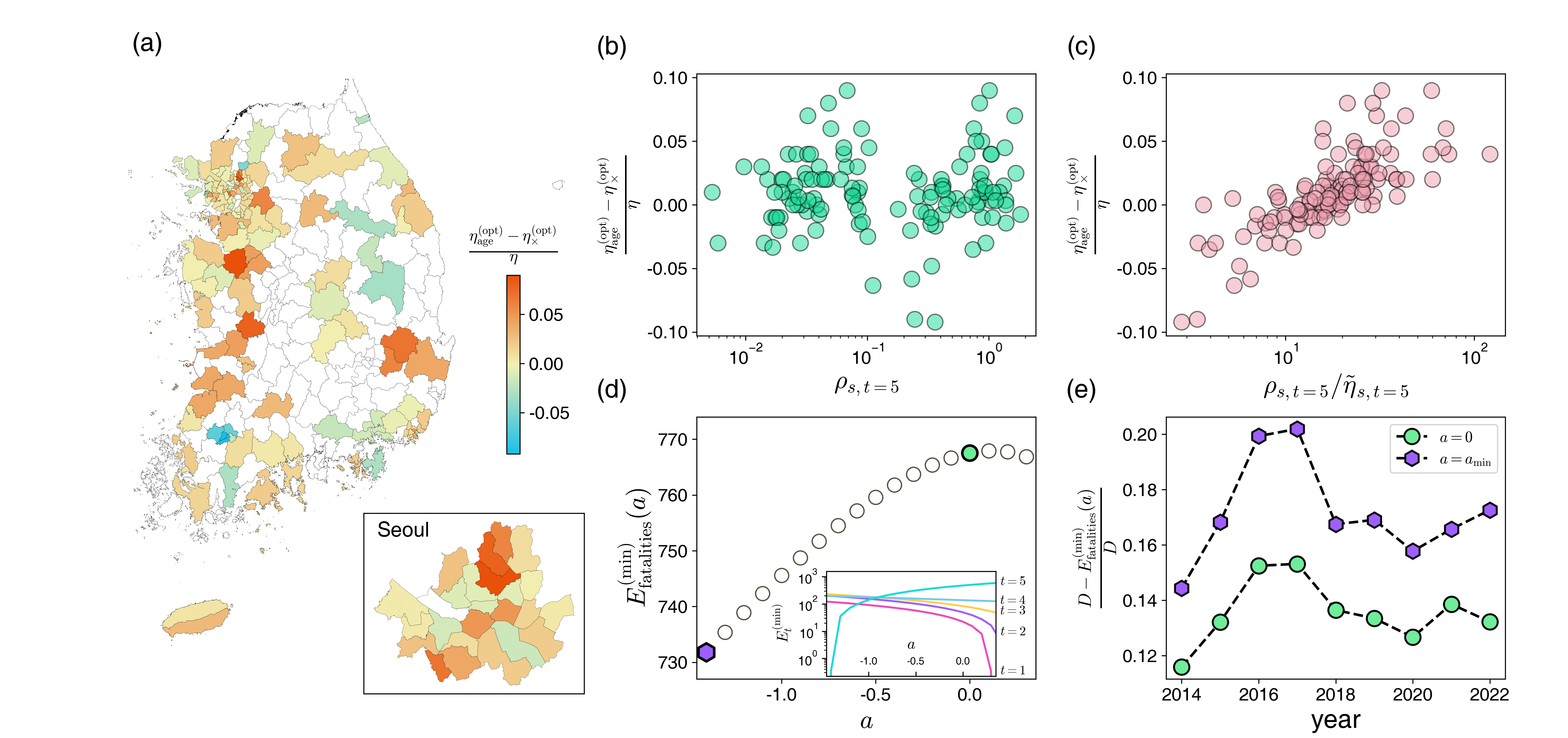}
\caption{
Comparison of age-aware and age-agnostic optimization outcomes.
(a) District-level values of the difference ${(\eta^{\rm (opt)}_{\rm age} - \eta^{\rm (opt)}_{\times})}/{\eta}$ across South Korea. The inset shows a magnified view of Seoul. Uncolored districts indicate no data satisfying the criteria. 
(b-c) Comparison of the difference ${(\eta^{\rm (opt)}_{\rm age} - \eta^{\rm (opt)}_{\times})}/{\eta}$ for $\rho_{s,5}$ (b) and $\rho_{s,5}/\tilde \eta_{s,5}$ (c). No distinct pattern is observed with $\rho_{s, 5}$, whereas a steady, roughly linear increase is seen with $\rho_{s,5}/\tilde \eta_{s,5}$ despite the log-scaled x-axis. (d) The minimized objective $E^{\rm (min)}_{\rm{fatalities}}$ as a function of the age-weighting parameter $a$ in 2022 is shown as a representative case, with $a_{\rm min}=-1.4$ and $a_{\rm max}=0.3$. The purple hexagon and green circle represent the cases of $a=a_{\rm min}$ and $a=0$, respectively. The inset shows the age-group contributions $E_t^{\rm (min)}$ to the minimized objective as functions of $a$. (e) The yearly trend of the reduction in $E^{\rm (min)}_{\rm{fatalities}}$ relative to the empirical reference value $D$ when $a=a_{\rm min}$ and $a=0$. A larger value indicates a greater reduction in the objective relative to $D$. Across all analyzed years, the objective reduction is greater for $a=a_{\rm min}$ than for $a=0$. 
}
\label{fig:E_a}
\end{figure}

To understand the effect of accounting for age-group distributions on the optimization process, we compare the age-aware optimization results to the age-agnostic baseline studied in the previous work~\cite{lee2020optimizing}, which uses the objective function in Eq.~(\ref{eq:energy_previous}). We first examine the relative change in hospital density compared to the original empirical density $\eta$, specifically ${\eta^{\rm (opt)}_{\rm age}}/{\eta}$ and ${\eta^{\rm (opt)}_{\times}}/{\eta}$, which are obtained by minimizing the objective function $E_{\rm fatalities}$ in Eq.~(\ref{eq:energy_st}) for $\eta^{\rm (opt)}_{\rm age}$ and $E(\vec{\eta})$ in Eq.~(\ref{eq:energy_previous}) for $\eta^{\rm (opt)}_{\times}$, respectively. Here, the subscript $\times$ represents the optimal hospital density obtained without accounting for age-group differences. Since elderly individuals account for a substantial portion of TB patients and fatalities, we analyze the optimization results with respect to the areal patient density of the oldest age group ($t = 5$, ages 80 and above), defined as $\rho_{s,5} = N_{s,5} / A_s$. The relative changes ${\eta^{\rm (opt)}_{\rm age}}/{\eta}$ and ${\eta^{\rm (opt)}_{\times}}/{\eta}$ are nearly identical [see the color map in Fig.~\ref{fig:E_a}(a) and the y-axis scales in Figs.~\ref{fig:E_a}(b) and~\ref{fig:E_a}(c)].

A higher density of elderly patients alone does not necessarily lead to a greater hospital allocation in the age-aware optimization [Fig.~\ref{fig:E_a}(b)], indicating that absolute patient density is not a sufficient predictor of post-optimization hospital gain. In line with Ref.~\cite{lee2020optimizing}, which showed that hospital gain or loss is governed by the characteristic patient density, we therefore examine the optimization outcomes from the perspective of the rescaled density $\rho_{s,5} / \tilde{\eta}_{s,5}$. We find that the rescaled patient density plays a dual role. First, the relative change in hospital density ${\eta^{\rm (opt)}_{\rm age}}/{\eta}$ exhibits a clear dependence on ${\rho_{s,5}}/{\tilde{\eta}_{s,5}}$, indicating that this quantity is associated with whether a district gains or loses hospitals after optimization [Supplementary Fig.~S1]. Second, the difference between the age-aware and age-agnostic optimizations, ${(\eta^{\rm (opt)}_{\rm age} - \eta^{\rm (opt)}_{\times})}/{\eta}$, also shows a systematic dependence on the same rescaled density [Fig.~\ref{fig:E_a}(c)]. Thus, the rescaled patient density ${\rho_{s,5}}/{\tilde{\eta}_{s,5}}$ is associated with both the direction of hospital redistribution and the magnitude of the difference between the age-aware and age-agnostic optima. 

The characteristic scale $\tilde{\eta}_{s,t}$ defined in Eq.~(\ref{eq:tilde_st}) is strongly correlated with the age-aggregated scale $\tilde{\eta}_s$, since the empirical fatality rate $\phi_{s,t}$ is constrained by the overall rate $\phi_s$. This correlation explains why the behavior of the hospital density difference also mirrors a similar dependence on the age-aggregated ratio ${\rho}/{\tilde{\eta}}$ [see Supplementary Fig.~S2].

\subsection*{Effect of age weighting on the fatality-minimization objective}

The age distribution of TB patients may vary over time due to demographic changes such as population aging. Rather than modeling such changes explicitly, we examine the sensitivity of the optimization result to the relative contributions of different age groups by introducing a simple linear weight of the form $at+b$ in Eq.~(\ref{eq:energy_st}). Incorporating this weighting, the objective function for optimization is redefined as follows:
\begin{equation}
E_{\rm {fatalities}}(\vec{\eta}; a) = \sum_{s,t} E_{s, t}= \sum_{s,t} (a t + b) N_{s,t} \exp\left(-{\eta_s}/{\tilde{\eta}_{s,t}}\right),
\label{eq:energy_st_fin}
\end{equation}
which allows us to express the minimized objective $E^{\rm (min)}_{\rm {fatalities}}$ in Eq.~(\ref{eq:energy_min}) as a function of $a$, such that $E^{\rm (min)}_{\rm {fatalities}}\equiv E(\vec{\eta}^{\rm (opt)}; a)$. We will show later that $b$ is not a free parameter but is determined by $a$. Positive values of $a$ assign greater relative weight to older age groups, whereas negative values assign greater relative weight to younger age groups. When $a = 0$, the age-weighted objective function $E_{\rm fatalities}(\vec{\eta}; a=0)$ in Eq.~(\ref{eq:energy_st_fin}) reduces to the original, unweighted formulation in Eq.~(\ref{eq:energy_st}). The slope $a$ in the age-weighting function is constrained to lie within a bounded range, $a_{\min} \leq a \leq a_{\max}$, to ensure non-negative weights for all age groups. The bounds are defined as follows:
\begin{align}
a_{\min} &= \min [ a \in \mathbb{R} \mid at + b \geq 0 \text{ for all } t \in \{1, 2, 3, 4, 5\} ], \nonumber \\
a_{\max} &= \max [ a \in \mathbb{R} \mid at + b \geq 0 \text{ for all } t \in \{1, 2, 3, 4, 5\} ], \nonumber
\end{align}
where $\mathbb{R}$ is the set of real numbers. The minimized objective $E^{\rm (min)}_{\rm{fatalities}}$ can be obtained as a function of $a$ for a given year, and the feasible range of $a$ is approximately $-1.4 \lesssim a \lesssim 0.3$ over the nine-year period, as shown in Supplementary Fig.~S3.

Since the optimization process begins from the empirical number of total fatalities, computed as $D = \sum_{s,t} D_{s,t}$ and listed in Table~\ref{tab:data_info}, the parameter $b$ in the age-weighting function is chosen to ensure that $E_{\rm{fatalities}}$ evaluated at the empirical $\vec{\eta}$ in Eq.~(\ref{eq:energy_st_fin}) equals the empirical value $D$ for a given $a$. That is, $\sum_{s, t}(at+b)D_{s,t}=D$ and this condition determines $b$ as a function of $a$:
\begin{equation}
b(a) = 1 - a \sum\limits_{s,t} t\, \frac{D_{s,t}}{D}.
\label{eq:b}
\end{equation}
All quantities on the right-hand side of Eq.~(\ref{eq:b}), except for $a$, are taken from the original empirical values prior to optimization. The characteristic hospital density is not altered by the age weighting because the fatality rate $\phi_{s,t} = {D_{s,t}}/{N_{s,t}}$ entering Eq.~(\ref{eq:tilde_st}) remains unchanged.

The behavior of the minimized objective $E^{\rm (min)}_{\rm{fatalities}}$ as a function of the age-weighting parameter $a$ is shown in Fig.~\ref{fig:E_a}(d), using the year 2022 as a representative case. The curve exhibits a slightly concave shape with a peak near $a = 0$, which is a typical pattern observed across all years. The respective age-group contributions $E_{t}^{\rm (min)}$ to the minimized objective, satisfying $E_{{\rm fatalities}}^{\rm (min)}=\sum_{t=1}^{5}E_{t}^{\rm (min)}$, show different behaviors as functions of $a$. The contribution $E^{\rm (min)}_5$ from the oldest group decreases rapidly with decreasing $a$ for $a<0$, while those of the other groups increase, as seen in the inset of Fig.~\ref{fig:E_a}(d). Among the age-group contributions, $E^{\rm (min)}_5$ shows the strongest variation with $a$.
The minimum value is $E^{\rm (min)}_{\rm{fatalities}} = 731.80$, corresponding to an approximately 17\% reduction in the weighted objective relative to its value for the empirical configuration ($D = 886$), at $a = -1.4 \simeq a_{\rm min}$, whereas the corresponding value at $a = 0$ is $767.54$ (about a 13\% reduction). The tendency $E^{\rm (min)}_{\rm{fatalities}}(a=0) > E^{\rm (min)}_{\rm{fatalities}}(a\simeq a_{\rm min})$ is consistently observed across all periods [Fig.~\ref{fig:E_a}(e)]. The curve $E^{\rm (min)}_{\rm{fatalities}}$ as a function of $a$ for all periods is displayed in Supplementary Fig.~S3.

\begin{figure}[b!]
\centering
\includegraphics[width=0.75\columnwidth]{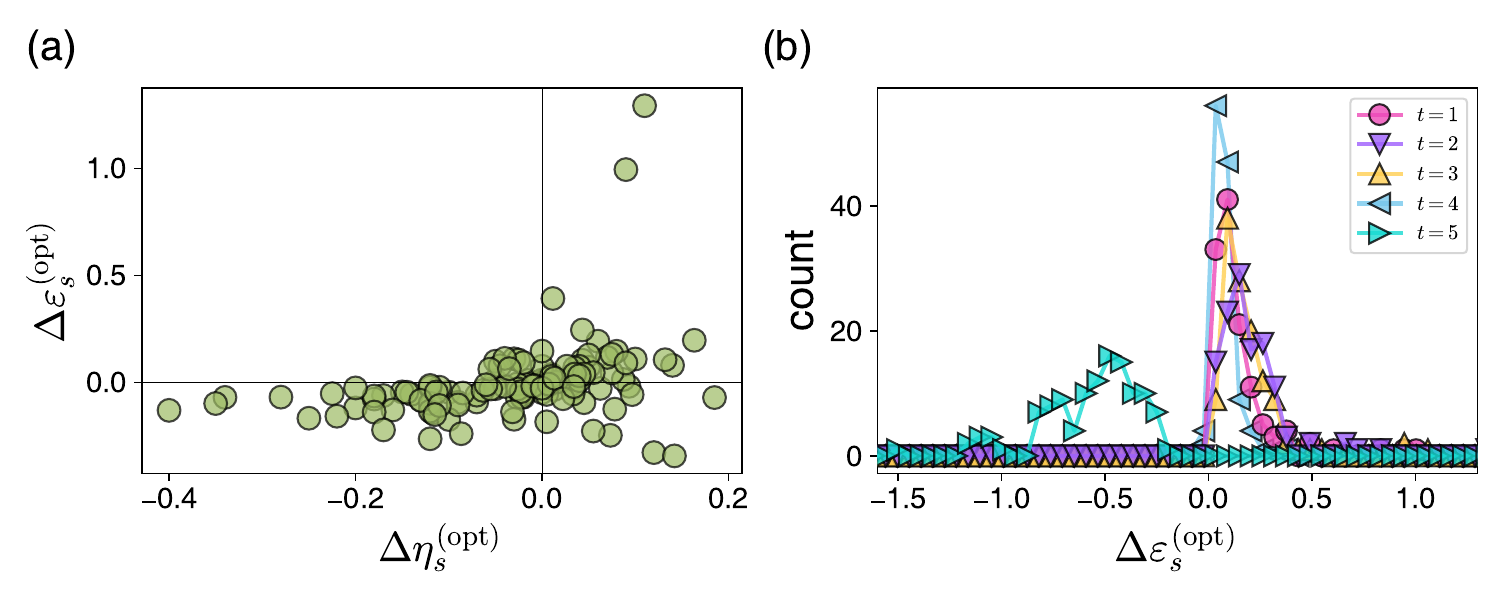}
\caption{District-level effects of age weighting on optimized allocation and objective contributions. (a) Relationship between the relative change in optimal hospital allocation $\Delta \eta^{\rm (opt)}_{s}$ and the relative change in the optimized fatality objective $\Delta \epsilon^{\rm (opt)}_{s}$ across districts for the $a\simeq a_\mathrm{min}$ case compared with the baseline case ($a=0$). (b) Distributions of the age-group-specific contributions $\Delta \epsilon^{\rm (opt)}_{s,t}$ across districts for each age group $t$. The contribution of the oldest age group ($t=5$, ages 80 and above) is concentrated in the negative range, unlike those of the younger age groups.}
\label{fig:diff_a}
\end{figure} 

To explore how age weighting influences the reallocation of hospitals and the optimization outcome at the district level, we compare two representative cases: the unweighted baseline with $a=0$ and a younger-age-weighted case with $a\simeq a_{\rm min}$. In both cases, the total number of hospitals remains constant, but the optimal allocation patterns differ due to the application of age-dependent weights in the objective function. For each district $s$, we define the relative changes in hospital allocation and the optimized objective as follows:
\begin{align}
    \Delta \eta_s^{\rm (opt)}\equiv \frac{\eta_s^{\rm (opt)}(a\simeq a_{\rm min})}{\eta_s}-\frac{\eta_s^{\rm (opt)}(a=0)}{\eta_s},~~~~~
    \Delta \epsilon_s^{\rm (opt)}\equiv \frac{E_s^{\rm (opt)}(a\simeq a_{\rm min})}{D_s}-\frac{E_s^{\rm (opt)}(a=0)}{D_s}. \nonumber
\end{align}
These quantities represent, respectively, the relative changes in hospital allocation and the district-level optimized fatality objective for each district when moving from the baseline to the younger-age-weighted case. Here, $a\simeq a_{\rm min}$ represents the younger-age-weighted extreme, while the opposite extreme, $a\simeq a_{\rm max}$, assigns greater relative weight to older age groups and is analyzed separately in Supplementary Fig.~S4.

Figure~\ref{fig:diff_a}(a) shows a scatter plot of $\Delta\eta_s^{\rm (opt)}$ versus $\Delta\epsilon_s^{\rm (opt)}$ for all districts. In the younger-age-weighted case, some districts lose hospital resources while the optimized objective decreases, whereas others gain resources while the objective increases. To clarify the origin of these different behaviors, we decompose the district-level objective change into age-group-specific contributions,
\begin{equation}
\Delta \epsilon_{s,t}^{\rm (opt)}
\equiv
\frac{E_{s,t}^{\rm (opt)}(a\simeq a_{\rm min})}{D_s}
-
\frac{E_{s,t}^{\rm (opt)}(a=0)}{D_s},
\qquad
\Delta \epsilon_s^{\rm (opt)}=\sum_t \Delta \epsilon_{s,t}^{\rm (opt)},
\nonumber
\end{equation}
where $E_{s,t}^{\rm (opt)}(a)$ denotes the age-group term in Eq.~(\ref{eq:energy_st_fin}) evaluated at the optimized configuration for $a$, for $t=1,\dots,5$ [Fig.~\ref{fig:diff_a}(b)].
This decomposition reveals a strong asymmetry across age groups. 
The contribution from the oldest group ($t=5$, ages 80 and above) is consistently negative across districts, whereas younger age groups typically show small positive contributions. 
Consequently, the net objective change in a district is determined by the balance between the negative contribution from the oldest group and the cumulative positive contributions from younger groups.

This composition helps explain the outcomes observed under age-weighted optimization. In the younger-age-weighted case, some districts receive more hospitals than in the baseline case and still exhibit an increase in the weighted fatality objective. In these districts, the positive contributions from younger age groups outweigh the negative contribution from the oldest group. The opposite behavior can be understood in the same manner. These observations indicate that age weighting modifies not only the spatial allocation of hospitals but also how allocation changes correspond to changes in the weighted objective. The comparison between age-weighted and baseline optimizations further shows that age weighting can alter spatial allocation patterns under fixed infrastructural constraints (as encoded in $\tilde{\eta}_{s,t}$). Rather than responding uniformly to hospital gains or losses, district-level outcomes reflect countervailing contributions from the oldest and younger age groups. These results highlight the value of examining age-resolved contributions when interpreting changes in the aggregate optimization outcome.

\section*{Discussion}
\label{sec:discussion}

Using the reconstructed age-resolved district-level TB data, we examine how age-dependent vulnerability affects the spatial allocation of hospitals in South Korea. The data are constructed by integrating age-stratified `si' and `do' (province) level spatial data with non-age-stratified `si', `gun', and `gu' (district) level spatial data, enabling optimization analyses that account for both demographic vulnerability and spatial variation at the district scale. Building on the existing hospital-allocation framework, we compare optimization results obtained with and without age-group differences. The relative changes in hospital density are nearly identical overall, but systematic district-level differences remain between the two optima. In particular, the rescaled patient density of the oldest age group, $\rho_{s,5}/\tilde{\eta}_{s,5}$, is systematically related to both the direction of hospital redistribution and the difference between the age-aware and age-agnostic allocations. These results suggest that incorporating age-specific information can reveal local differences that are not apparent from the overall allocation patterns alone.

The age-weighting analysis further allows us to examine how contributions from different age groups affect the optimization result. As the age-weighting parameter is varied, the minimized objective shows different responses across age groups, with the oldest age group contributing most strongly to its variation. In the comparison between $a\simeq a_{\rm min}$ and the baseline case $a=0$, the contribution from the oldest age group is consistently negative across districts, whereas younger age groups typically make smaller positive contributions. Consequently, the district-level change reflects the balance of these countervailing contributions. Some districts receive more hospital resources than in the baseline case while showing an increase in the weighted fatality objective, whereas the opposite behavior is also observed. This analysis helps clarify how changes in hospital allocation and in the aggregate optimization outcome arise from different age-group contributions.

Several limitations of this study should be stated. Age-disaggregated TB data at the district level are reconstructed from province-level distributions, which implicitly assumes homogeneous age composition within each province and may obscure finer-scale demographic variation. In addition, hospital accessibility is represented through areal hospital density and does not explicitly account for differences in capacity, quality of care, or transportation infrastructure. The linear age-weighting scheme is intended as a simple means of examining changes in the relative contributions of different age groups and should not be interpreted as a direct demographic forecast. As a result, the optimized configurations should be interpreted as relative allocation patterns within the present model rather than as direct prescriptions for policy implementation.

Despite these limitations, the framework provides a consistent basis for comparative or exploratory studies of demographic effects in spatial healthcare allocation. Because the objective function is explicitly formulated as a weighted sum over demographic groups, the approach may be adapted to other diseases or risk profiles by modifying the weighting scheme or the definition of fatality risk, provided that appropriate data and disease-specific validation are available. In this sense, the present framework is intended primarily for examining how demographic heterogeneity can affect spatial allocation patterns rather than as a prescriptive tool for policy design.
\section*{Data Availability}
The dataset supporting this study has been deposited in the Dryad Digital Repository and is available for peer review at the following private link: \href{https://datadryad.org/share/qfxSkgpvSxiVsZhnUP8jLxwMMIXUiLwCaVrP5Y0Djek}{https://datadryad.org/share/qfxSkgpvSxiVsZhnUP8jLxwMMIXUiLwCaVrP5Y0Djek}. This dataset contains annual tuberculosis (TB) statistics across administrative districts in South Korea from 2014 to 2022. It includes district-level counts of hospitals, TB patients, and fatalities, along with age-stratified proportions of new cases and deaths (by 10-year age groups from 0–9 to 80+). 
The dataset was reconstructed using publicly available provincial-level data from the Korea Disease Control and Prevention Agency (KDCA) and the Korean Statistical Information Service (KOSIS), and harmonized with higher-resolution regional information to produce the reconstructed age-resolved district-level TB dataset used in this study. All variables are fully documented in the accompanying metadata and README file. The dataset will be assigned a DOI and made publicly available upon publication.

\section*{Code Availability}
The code used for data preprocessing, age-stratified fatality modeling, and hospital optimization is openly available at \href{https://github.com/kwyosu7/Tuberculosis_hospital_distribution_optimization}{https://github.com/kwyosu7/Tuberculosis\_hospital\_distribution\_optimization}. The repository includes Python scripts for constructing the dataset, implementing optimization routines, and generating the figures used in the manuscript.

\section*{Funding} 
This work was supported by the National Research Foundation of Korea (NRF) through Grants RS-2026-25488703 (S.-W.S.) and RS-2024-00341317 (M.J.L.), and by KIAS Individual Grant No. CG079902 (D.-S.L.). 

\section*{Acknowledgments} 
We thank APCTP, Pohang, Korea, for its hospitality during the Topical Research Program [APCTP-2025-T04], from which this work benefited greatly.

\bibliography{ref}

\begin{thebibliography}{10}
\urlstyle{rm}
\expandafter\ifx\csname url\endcsname\relax
  \def\url#1{\texttt{#1}}\fi
\expandafter\ifx\csname urlprefix\endcsname\relax\def\urlprefix{URL }\fi
\expandafter\ifx\csname doiprefix\endcsname\relax\def\doiprefix{DOI: }\fi
\providecommand{\bibinfo}[2]{#2}
\providecommand{\eprint}[2][]{\url{#2}}

\bibitem{WHO2023}
\bibinfo{author}{(WHO), W. H.~O.}
\newblock \bibinfo{title}{Tuberculosis data}.
\newblock \bibinfo{howpublished}{\url{https://www.who.int/tb/data/en/}} (\bibinfo{year}{2023}).
\newblock \bibinfo{note}{Accessed: 2025-04-14}.

\bibitem{glaziou2015global}
\bibinfo{author}{Glaziou, P.}, \bibinfo{author}{Sismanidis, C.}, \bibinfo{author}{Floyd, K.} \& \bibinfo{author}{Raviglione, M.}
\newblock \bibinfo{journal}{\bibinfo{title}{Global epidemiology of tuberculosis}}.
\newblock {\emph{\JournalTitle{Cold Spring Harbor perspectives in medicine}}} \textbf{\bibinfo{volume}{5}}, \bibinfo{pages}{a017798} (\bibinfo{year}{2015}).

\bibitem{renner2024hospitals}
\bibinfo{author}{Renner, A.-T.}
\newblock \bibinfo{title}{Hospitals as social infrastructure: accessible for all?}
\newblock In \emph{\bibinfo{booktitle}{Handbook of Social Infrastructure}}, \bibinfo{pages}{20--38} (\bibinfo{publisher}{Edward Elgar Publishing}, \bibinfo{year}{2024}).

\bibitem{chung2024access}
\bibinfo{author}{Chung, S.} \emph{et~al.}
\newblock \bibinfo{journal}{\bibinfo{title}{Access to emergency services: A new york city case study}}.
\newblock {\emph{\JournalTitle{Transportation Research Interdisciplinary Perspectives}}} \textbf{\bibinfo{volume}{25}}, \bibinfo{pages}{101111} (\bibinfo{year}{2024}).

\bibitem{lee2025global}
\bibinfo{author}{Lee, H.}, \bibinfo{author}{Kim, J.}, \bibinfo{author}{Kim, J.} \& \bibinfo{author}{Park, Y.-J.}
\newblock \bibinfo{journal}{\bibinfo{title}{Review of the global burden of tuberculosis in 2023: Insights from the who global tuberculosis report 2024}}.
\newblock {\emph{\JournalTitle{Public Health Weekly Report}}} \textbf{\bibinfo{volume}{18}}, \bibinfo{pages}{S55--S69} (\bibinfo{year}{2025}).

\bibitem{cdc2013tbcurriculum}
\bibinfo{author}{{Centers for Disease Control and Prevention}}.
\newblock \emph{\bibinfo{title}{Core Curriculum on Tuberculosis: What the Clinician Should Know}} (\bibinfo{publisher}{Centers for Disease Control and Prevention}, \bibinfo{address}{Atlanta, GA}, \bibinfo{year}{2013}).

\bibitem{hopewell2006international}
\bibinfo{author}{Hopewell, P.~C.}, \bibinfo{author}{Pai, M.}, \bibinfo{author}{Maher, D.}, \bibinfo{author}{Uplekar, M.} \& \bibinfo{author}{Raviglione, M.~C.}
\newblock \bibinfo{journal}{\bibinfo{title}{International standards for tuberculosis care}}.
\newblock {\emph{\JournalTitle{The Lancet infectious diseases}}} \textbf{\bibinfo{volume}{6}}, \bibinfo{pages}{710--725} (\bibinfo{year}{2006}).

\bibitem{KDCA}
\bibinfo{note}{Korea Disease Control and Prevention Agenc. Available from \url{https://www.kdca.go.kr} (Accessed: 12.15.2023)}.

\bibitem{KOSIS}
\bibinfo{note}{Korean Statistical Information Service. Available from \url{http://kosis.kr} (Accessed: 12.15.2023)}.

\bibitem{um2009scaling}
\bibinfo{author}{Um, J.}, \bibinfo{author}{Son, S.-W.}, \bibinfo{author}{Lee, S.-I.}, \bibinfo{author}{Jeong, H.} \& \bibinfo{author}{Kim, B.~J.}
\newblock \bibinfo{journal}{\bibinfo{title}{Scaling laws between population and facility densities}}.
\newblock {\emph{\JournalTitle{Proceedings of the National Academy of Sciences}}} \textbf{\bibinfo{volume}{106}}, \bibinfo{pages}{14236--14240} (\bibinfo{year}{2009}).

\bibitem{stephan1977territorial}
\bibinfo{author}{Stephan, G.~E.}
\newblock \bibinfo{journal}{\bibinfo{title}{Territorial division: The least-time constraint behind the formation of subnational boundaries}}.
\newblock {\emph{\JournalTitle{Science}}} \textbf{\bibinfo{volume}{196}}, \bibinfo{pages}{523--524} (\bibinfo{year}{1977}).

\bibitem{gusein1982bunge}
\bibinfo{author}{Gusein-Zade, S.~M.}
\newblock \bibinfo{journal}{\bibinfo{title}{Bunge's problem in central place theory and its generalizations}}.
\newblock {\emph{\JournalTitle{Geographical Analysis}}} \textbf{\bibinfo{volume}{14}}, \bibinfo{pages}{246--252} (\bibinfo{year}{1982}).

\bibitem{gastner2006optimal}
\bibinfo{author}{Gastner, M.~T.} \& \bibinfo{author}{Newman, M.~E.}
\newblock \bibinfo{journal}{\bibinfo{title}{Optimal design of spatial distribution networks}}.
\newblock {\emph{\JournalTitle{Physical Review E}}} \textbf{\bibinfo{volume}{74}}, \bibinfo{pages}{016117} (\bibinfo{year}{2006}).

\bibitem{kim2012internet}
\bibinfo{author}{Kim, D.}, \bibinfo{author}{Son, S.-W.} \& \bibinfo{author}{Jeong, H.}
\newblock \bibinfo{journal}{\bibinfo{title}{Demographic studies of internet routers}}.
\newblock {\emph{\JournalTitle{Journal of the Korean Physical Society}}} \textbf{\bibinfo{volume}{60}}, \bibinfo{pages}{585--589} (\bibinfo{year}{2012}).

\bibitem{lee2017spatial}
\bibinfo{author}{Lee, M.~J.} \& \bibinfo{author}{Kim, B.~J.}
\newblock \bibinfo{journal}{\bibinfo{title}{Spatial uniformity in the power-grid system}}.
\newblock {\emph{\JournalTitle{Physical Review E}}} \textbf{\bibinfo{volume}{95}}, \bibinfo{pages}{042316} (\bibinfo{year}{2017}).

\bibitem{wuellner2010resilience}
\bibinfo{author}{Wuellner, D.~R.}, \bibinfo{author}{Roy, S.} \& \bibinfo{author}{D’Souza, R.~M.}
\newblock \bibinfo{journal}{\bibinfo{title}{Resilience and rewiring of the passenger airline networks in the united states}}.
\newblock {\emph{\JournalTitle{Physical Review E}}} \textbf{\bibinfo{volume}{82}}, \bibinfo{pages}{056101} (\bibinfo{year}{2010}).

\bibitem{lee2022spatial}
\bibinfo{author}{Lee, J.-H.}, \bibinfo{author}{Jo, J.}, \bibinfo{author}{Kim, J.~W.}, \bibinfo{author}{Lee, K.} \& \bibinfo{author}{Choi, M.~Y.}
\newblock \bibinfo{journal}{\bibinfo{title}{Spatial distributions of restaurants emerging from pedestrian behavior and online information sharing}}.
\newblock {\emph{\JournalTitle{Physica A: Statistical Mechanics and its Applications}}} \textbf{\bibinfo{volume}{597}}, \bibinfo{pages}{127265} (\bibinfo{year}{2022}).

\bibitem{kwon2025quantifying}
\bibinfo{author}{Kwon, Y.}, \bibinfo{author}{Lee, M.~J.} \& \bibinfo{author}{Son, S.-W.}
\newblock \bibinfo{journal}{\bibinfo{title}{Quantifying traffic patterns with percolation theory: a case study of seoul roads}}.
\newblock {\emph{\JournalTitle{Journal of the Korean Physical Society}}} \textbf{\bibinfo{volume}{86}}, \bibinfo{pages}{693--700} (\bibinfo{year}{2025}).

\bibitem{lee2020optimizing}
\bibinfo{author}{Lee, M.~J.}, \bibinfo{author}{Kim, K.}, \bibinfo{author}{Son, J.} \& \bibinfo{author}{Lee, D.-S.}
\newblock \bibinfo{journal}{\bibinfo{title}{Optimizing hospital distribution across districts to reduce tuberculosis fatalities}}.
\newblock {\emph{\JournalTitle{Scientific Reports}}} \textbf{\bibinfo{volume}{10}}, \bibinfo{pages}{8603} (\bibinfo{year}{2020}).

\bibitem{hughes1996random}
\bibinfo{author}{Hughes, B.~D.}
\newblock \emph{\bibinfo{title}{Random walks and random environments}} (\bibinfo{publisher}{Oxford University Press}, \bibinfo{year}{1996}).

\bibitem{Rajagopalan2001}
\bibinfo{author}{Thomas, T.~Y.} \& \bibinfo{author}{Rajagopalan, S.}
\newblock \bibinfo{journal}{\bibinfo{title}{Tuberculosis and aging: A global health problem}}.
\newblock {\emph{\JournalTitle{Clinical Infectious Diseases}}} \textbf{\bibinfo{volume}{33}}, \bibinfo{pages}{1034--1039} (\bibinfo{year}{2001}).

\end{thebibliography}

\section*{Author Contributions Statement}
Yongsung Kwon, Deok-Sun Lee, Mi Jin Lee, and Seung-Woo Son designed and conceptualized the study and contributed to writing the manuscript. Yongsung Kwon and Mi Jin Lee collected and digitized the TB dataset. Deok-Sun Lee assisted with the early generation of the datasets. Yongsung Kwon developed and implemented the optimization models, conducted the computational analysis, and performed the spatial data processing and visualization. Deok-Sun Lee, Mi Jin Lee, and Seung-Woo Son supervised the project and provided critical guidance on the methodological framework. All authors reviewed and edited the manuscript.

\section*{Competing Interests} 
The authors declare no competing interests.

\clearpage

\section*{Supplemental Material}
\renewcommand{\thefigure}{S\arabic{figure}}   
\setcounter{figure}{0}                        
\begin{figure*}[h!]
\centering
\includegraphics[width=1\textwidth]{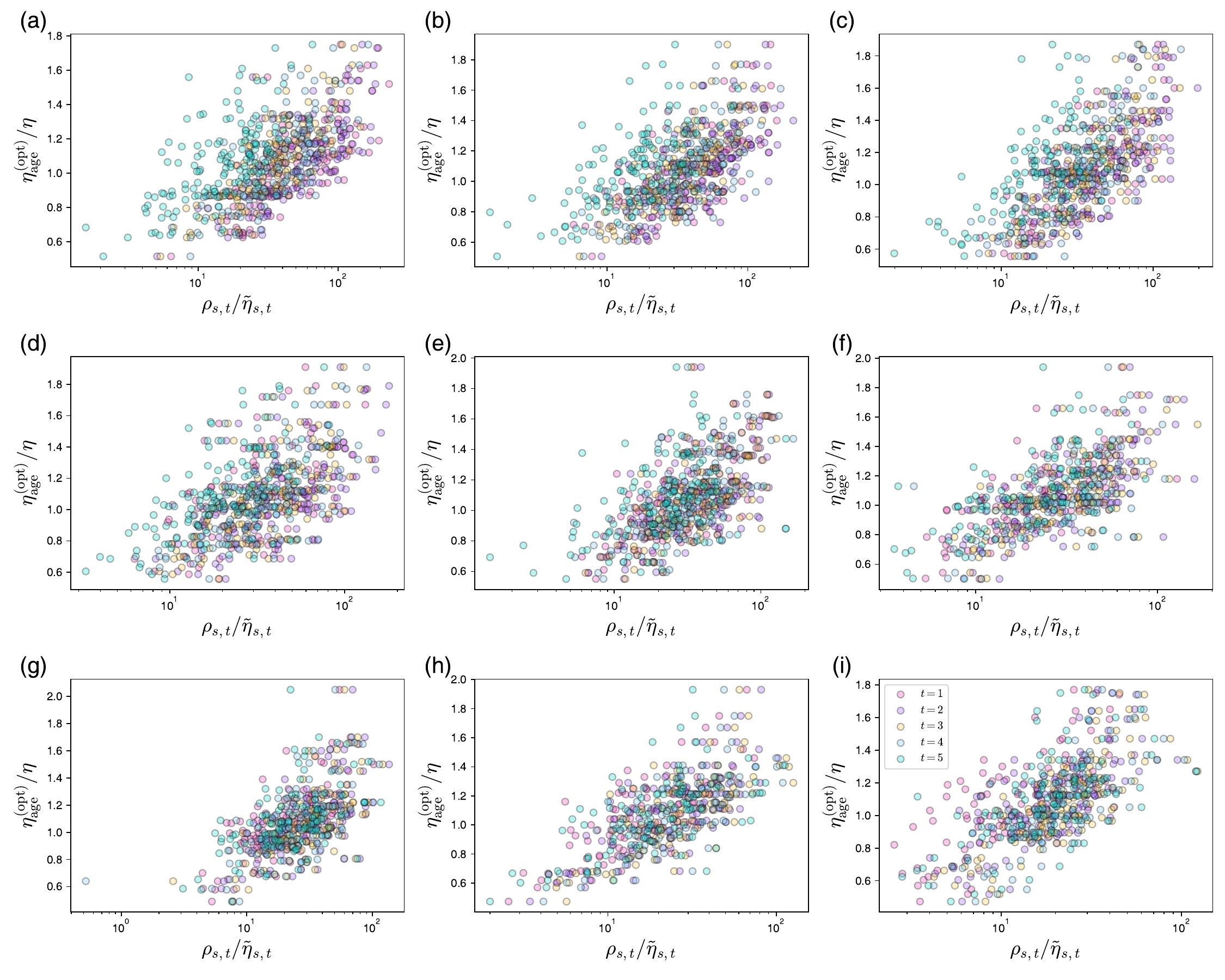}
\caption{
Plot of the ratio $\frac{\eta^{\rm (opt)}_{\rm age}}{\eta}$ versus the rescaled patient density $\frac{\rho_{s,t}}{\tilde{\eta}_{s,t}}$ from 2014 to 2022 (a-i). All age groups exhibit an increasing trend along the log-scaled x-axis. 
}
\label{suppfig:1}
\end{figure*}

\begin{figure*}
\centering
\includegraphics[width=1\textwidth]{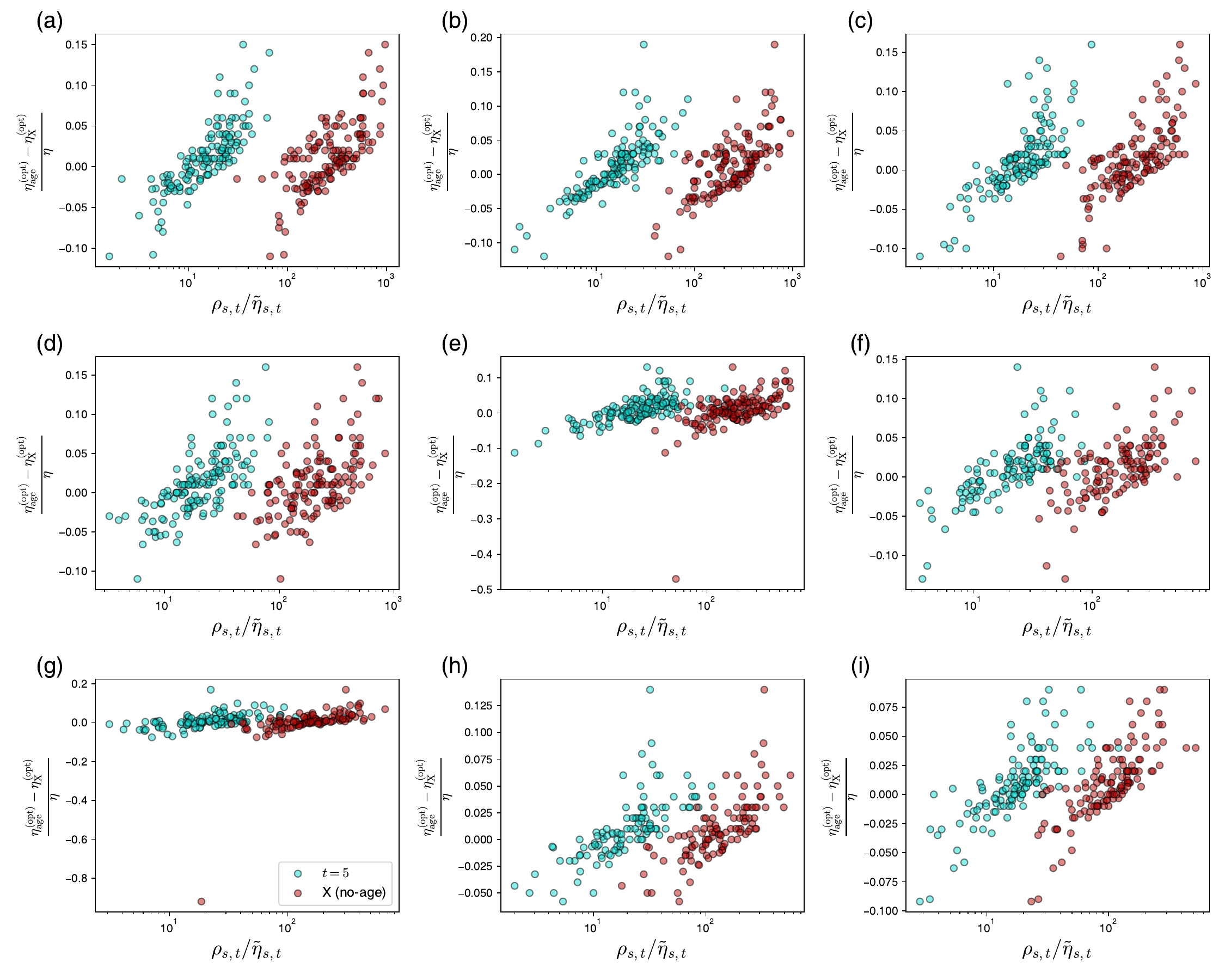}
\caption{ 
The difference between relative changes in hospital density $\frac{\eta^\mathrm{(opt)}_{\mathrm{age}} - \eta^\mathrm{(opt)}_{\times}}{\eta}$ versus the rescaled patient density $\frac{\rho_{s,t}}{\tilde{\eta}_{s,t}}$ from 2014 to 2022 (a-i) in $t=5$ age group and no-age case. 
}
\label{suppfig:2}
\end{figure*}

\begin{figure*}
\centering
\includegraphics[width=1\textwidth]{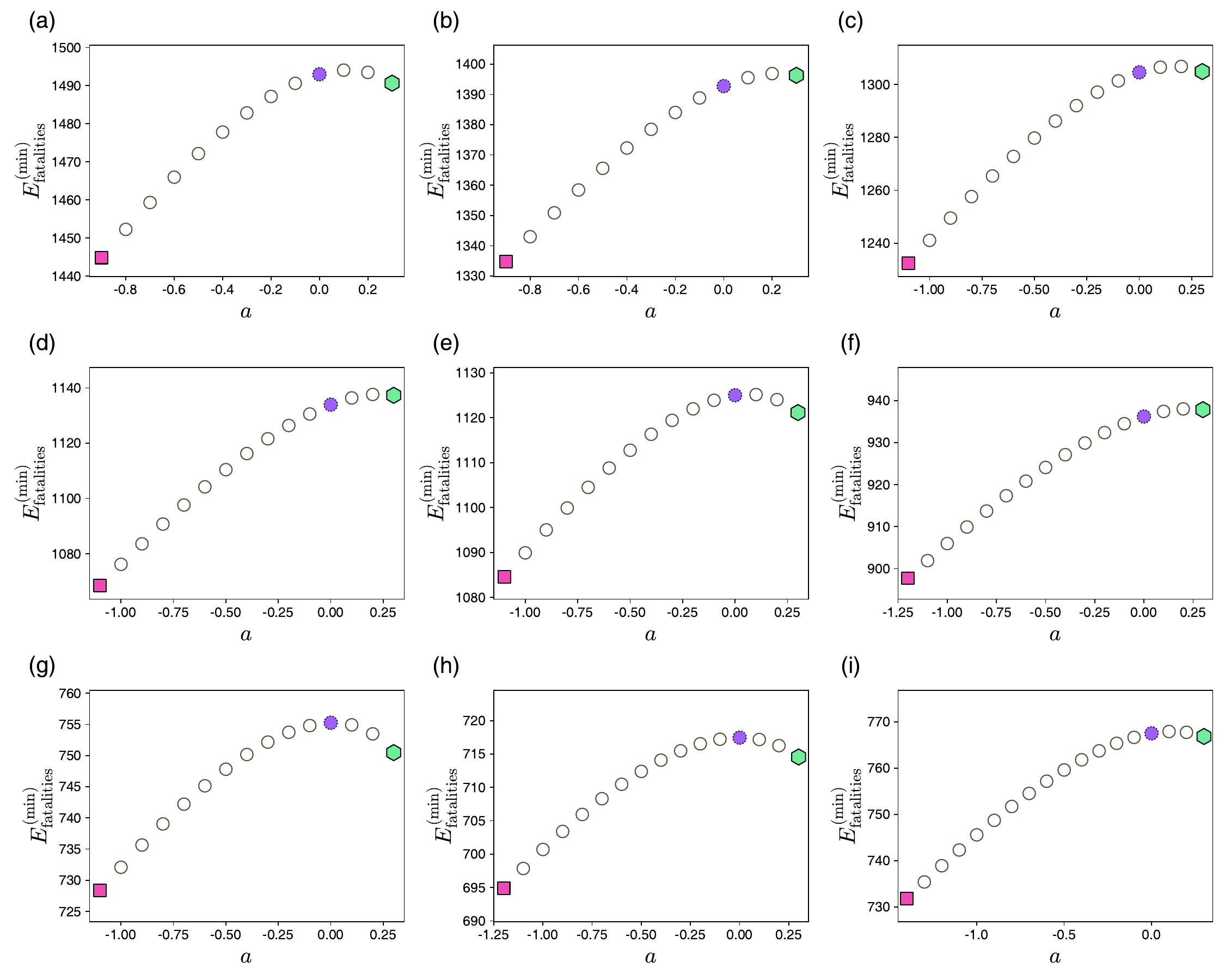}
\caption{
The minimum fatalities $E^{\rm (min)}_{\rm{fatalities}}$ as a function of the age weighting parameter $a$ from 2014 to 2022 is shown representatively, with $a_{\rm min}$ and $a_{\rm max}$. The pink square, green hexagon and purple circle represent the cases of $a=a_{\rm min}$, $a=a_{\rm max}$ and $a=0$, respectively.
}
\label{suppfig:3}
\end{figure*}

\begin{figure*}
\centering
\includegraphics[width=1\textwidth]{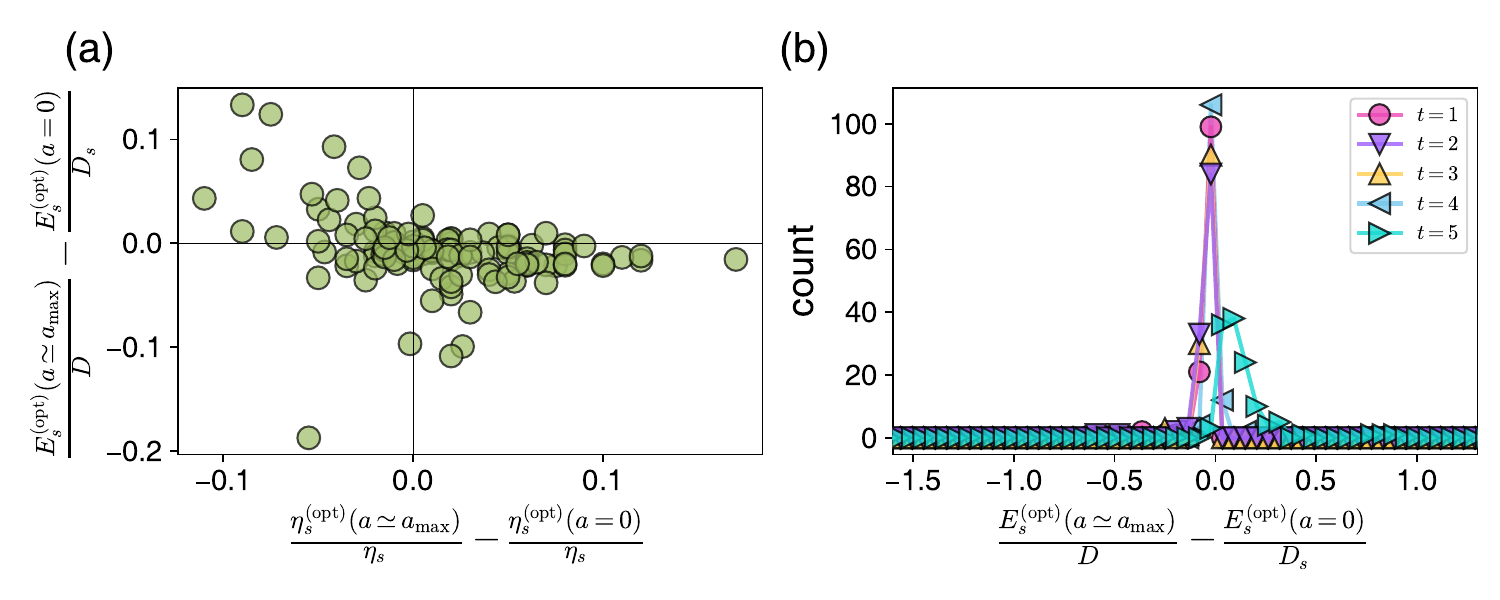}
\caption{ 
(a) Relationship between relative changes in optimal hospital allocation $\frac{\eta_s^{\rm (opt)}(a\simeq a_{\rm max})}{\eta_s}-\frac{\eta_s^{\rm (opt)}(a=0)}{\eta_s}$ versus fatality $\frac{E_s^{\rm (opt)}(a\simeq a_{\rm max})}{D}-\frac{E_s^{\rm (opt)}(a=0)}{D_s}$ across districts under the $a\simeq a_\mathrm{max}$ case compared to the baseline case ($a=0$). (b) A distribution of fatality change $\frac{E_s^{\rm (opt)}(a\simeq a_{\rm max})}{D}-\frac{E_s^{\rm (opt)}(a=0)}{D_s}$ for age group $t$. $\frac{E_s^{\rm (opt)}(a\simeq a_{\rm max})}{D}-\frac{E_s^{\rm (opt)}(a=0)}{D_s}$ for $t=5$ (ages 80 and above) concentrated in the negative range, unlike other age groups.
}
\label{suppfig:4}
\end{figure*}

\end{document}